\documentclass[12pt]{article}
\begin{document}

\title{Ghost Condensate Busting}


\author{
Neven Bili\'c$^1$\thanks{bilic@thphys.irb.hr}, Gary B.~Tupper$^2$\thanks{gary.tupper@uct.ac.za}
 and Raoul D.~Viollier$^2$\thanks{raoul.viollier@uct.ac.za}
 \\
 $^1$Rudjer Bo\v{s}kovi\'{c} Institute, 10002 Zagreb, Croatia \\
$^2$Centre of Theoretical Physics and Astrophysics,\\
University of Cape Town,  Rondebosch 7701, South Africa\\
}

\maketitle

\begin{abstract}
Applying the Thomas-Fermi approximation to renormalizable field
theories, we construct ghost condensation models that are free of
the instabilities associated with violations of the null-energy condition.
\end{abstract}




The null-energy condition (NEC), $T_{\mu \nu} n^{\mu} n^{\nu} \geq 0$,
where $T_{\mu \nu}$ is the energy-momentum tensor and $n_{\mu}$ an arbitrary 
light-like vector, is satisfied by all ordinary forms of interacting matter 
fields which are characterized by a positive kinetic energy density. For example, 
in the case of the canonical complex scalar field lagrangian
${\cal L} = |\partial \Phi|^{2} - U \left( |\Phi|^{2} \right)$, one obtains
$T_{\mu \nu} n^{\mu} n^{\nu} = 2 | n^{\mu} \Phi_{, \mu} |^{2} \geq 0$.
It is easy to imagine a violation of the NEC by reversing the sign of the kinetic 
energy density term, as in phantom models of dark energy \cite{cal1}. 
However, such theories 
invariably suffer from severe classical \cite{hsu2} and quantum  
\cite{bun3,cli4} instabilities.


Related ideas have been put forward under the heading of `ghost condensation' 
\cite{ark5}, where
one starts with a general lagrangian density 
${\cal L} = M^{4} P(X)$
that depends only
on the dimensionless kinetic energy density 
$X = g^{\mu \nu} \theta_{, \mu} \theta_{, \nu}$,
with $M$ being some mass scale. Here the function $P(X)$ expanded in powers of $X$ 
usually contains a
phantom or ghostlike linear term $- X/2$. The presence of higher-order powers 
of $X$ in $P(X)$ allows the field equation for $\theta$, i.e.
\begin{equation}
\nabla^{\mu}  \left[ P' (X)  \theta_{, \mu} \right] = 0 ,
\label{eq1}
\end{equation}
to have nontrivial solutions, obeying $P'(X_{c}) = 0$ for some value $X_{c}$. 
If $P''(X_{c}) > 0$, 
the condensate at $X_{c}$ is stable against small homogeneous fluctuations. 
One thus has a setup with a Lorentz breaking vacuum expectation value ($\dot{\theta} \neq 0$)
in a state with a stress-energy tensor of the vacuum form:
$T_{\mu}^{\nu} = - \delta_{\mu}^{\nu}\, P \left( X_{c} \right)$.
For example, the second degree polynomial
\begin{equation}
P(X) = (X - 1)^{2}/4
\label{eq2}
\end{equation}
fulfills both conditions at $X_{c} = 1$.

The canonical ghost condensate is still plagued by inconsistencies, however, as is 
evident from
\begin{equation}
T_{\mu \nu}  n^{\mu} n^{\nu} = 2  P' (X) (n^{\mu} \theta_{, \mu})^{2} \;\; .
\label{eq3}
\end{equation}
which violates the NEC for $X < X_{c}$.
In order to analyze this, we note that $\theta = ct$ with arbitrary $c$, 
is a solution of
(\ref{eq1}) in Minkowski space-time. Then, defining
$\theta = ct + \pi$,
one arrives at \cite{hsu2}
\begin{equation}
P(X) \simeq P(c^{2}) + \left[ 2 c^{2} P'' (c^{2}) + P'(c^{2}) \right] \dot{\pi}^{2} -
P' (c^{2}) \left( \vec{\nabla} \pi \right)^{2}  ,
\label{eq4}
\end{equation}
which exhibits a gradient instability for $P'(c^{2}) < 0$ violating the NEC.
Thus the ghost condensate is intrinsically unstable to inhomogeneous perturbations \cite{hsu2}.
 Indeed, Krotov {\it et al} \cite{kro6} have demonstarated that the simple polynomial (\ref{eq2}),
 as well as a much larger class of more sophisticated models, are unstable against 
the formation of holes in the ghost condensate, 
where NEC-violations occur in regions of the size $1/M$.

Higher derivatives that appear in  modified general relativity theories do not cure the instability
for the canonical ghost condensate or make it safe \cite{kal}.
In fact, it has been demonstrated \cite{ani} that the inclusion of higher derivatives 
would bring the system to the NEC violating region $X<X_c$.

As ghost condensation depends on higher powers of $X$, it can only be described by a
nonrenormalizable effective field theory that is valid up to some cutoff energy scale.
 In fact, 
as has been suggested in \cite{kro6}, 
the resolution of the hole instability may be expected in the theory's 
ultraviolet (UV) completion. Aside from some exotic constructs 
in string theory \cite{muk7}, Gasser {\it et al.} \cite{gas8} have attempted
 to produce ghost condensation via fermion loop corrections to the Higgs-Yukawa model.
 However, the validity of the latter model, which by construction proposes to
generate a
loop-induced NEC-violation, has been challenged by O'Connell \cite{con9}. 
It has even been suggested \cite{ada10} that UV completion may not be possible.

In this note, we want to show that by applying a well-defined approximation to 
renormalizable field theories, one may construct ghost condensation models
without producing NEC-violating holes in the ghost condensate. In other words, 
our models are free from NEC-violating instabilities, simply because their progenitor
field theories do not violate the NEC.

Consider for example the renormalizable Abelian $\sigma$ model
\begin{equation}
{\cal L}_{\sigma} = g^{\mu \nu} \Phi_{, \mu}^{*} \Phi_{, \nu} -
m^{2} |\Phi|^{2} - \lambda | \Phi |^{4}  .
\label{eq5}
\end{equation}
The complex scalar field $\Phi$ can be parameterized in terms of the real scalar 
fields, $\phi$ and $\theta$, as usual
\begin{equation}
\Phi = \frac{\phi}{\sqrt{2}}  {\rm e}^{i m \theta} \, ,
\label{eq6}
\end{equation}
yielding the lagrangian
\begin{equation}
{\cal L}_{\sigma} = \frac{1}{2}  g^{\mu \nu} \phi_{, \mu} \phi_{, \nu} +
\frac{m^{2}}{2}(X - 1)  \phi^{2} - 
\frac{\lambda}{4}  \phi^{4}  ,
\label{eq7}
\end{equation}
with 
\begin{equation}
X = g^{\mu \nu} {\theta}_{, \mu}
\theta_{, \nu}\, .
 \label{eq1104}
\end{equation}
 We now introduce the
Thomas-Fermi approximation \cite{par,bil4}
\begin{equation}
g^{\mu \nu} 
\frac{\phi_{, \mu}}{\phi} \: \frac{\phi_{, \nu}}{\phi} \ll m^{2}  ,
\label{eq8}
\end{equation}
restricting the wavelengths of the fields variations to much larger than 
the inverse mass scale, and yielding
\begin{equation}
{\cal L}_{\sigma \rm TF} = \frac{m^{2}}{2}  
(X - 1) \phi^{2} -  
\frac{\lambda}{4}  \phi^{4}  .
\label{eq9}
\end{equation}
Thus, equation (\ref{eq9}) is a valid long-wavelength approximation to the original 
renormalizable field theory given by (\ref{eq5}). 
The field equation for $\phi$ derived from ${\cal L}_{\sigma \rm TF}$ now 
becomes an algebraic equation
\begin{equation}
\phi \left[ m^{2} (X - 1) - \lambda \phi^{2} \right] = 0  ,
\label{eq10}
\end{equation}
with the solution
\begin{eqnarray}
\phi^{2} = \left\{ 
\begin{array}{rcl}
m^{2} (X - 1)/\lambda\, , &  \; \;\;\;\; & X > 1\\[.25cm]
                    0\, , &  \; \;\;\;\; & X \leq 1 .
\end{array} \right.
\label{eq11}
\end{eqnarray}
Note that, different from \cite{gas8}, we have $m^{2} > 0$ so we are not in the
broken symmetry phase. Instead, the nontrivial solution for $\phi$ is provided
 by the centrifugal barrier in its potential. Thus the Thomas-Fermi approximation 
is analogous to the approximations leading from the linear to the 
nonlinear $\sigma$-model.

Eliminating $\phi^{2}$ from ${\cal L}_{\sigma \rm TF}$ and defining
\begin{equation}
M^{4} = m^{4}/ \lambda  ,
\label{eq12}
\end{equation}
we obtain an {\em unholey} ghost condensate model with the
lagrangian
\begin{eqnarray}
{\cal L}_{\rm GC} =  M^{4} P(X) = M^{4} \left\{ 
\begin{array}{rcl}
(X - 1)^{2}/4\, , &  \;\;\;\;\;& X > 1\\[.25cm]
            0 \, ,&  \;\;\;\;\;& X \leq 1  .
\end{array} \right.
\label{eq13}
\end{eqnarray}
The key difference between (\ref{eq2}) and (\ref{eq13}), is that in the latter,
 the problematic region $X < 1$ does not exist. In fact, it has disappeared,
 together with the hole
instability of \cite{kro6}, and the would-be violation of the NEC. 
All this is a simple consequence of the initial model described in (\ref{eq5}) 
when subject to the long-wavelength Thomas-Fermi
approximation.

In Minkowski space-time any $\theta \propto t$ is a solution to (\ref{eq1}) and $\phi$
is a constant so the Thomas-Fermi approximation is exact. Hence, it is useful to consider a Friedmann-Robertson-Walker cosmology, in order to exhibit the self-consistency of the Thomas-Fermi approximation. In this case  $X = \dot{\theta}^{2} > 1$ holds, yielding
\begin{equation}
\phi^{2} = \frac{m^{2}}{\lambda} (X - 1) \, .
\label{eq038}
\end{equation}
The $\theta$ field equation (\ref{eq1}) thus becomes nontrivial,
\begin{equation}
\frac{1}{a^{3}}  \frac{d}{dt} \;
\left[ a^{3} (X - 1) \sqrt{X} \right] = 0  ,
\label{eq039}
\end{equation}
where $a(t)$ is the scale factor. Using (\ref{eq038}) and (\ref{eq039}), we find
\begin{equation}
\left| \frac{\dot{\phi}}{\phi} \right| =
\left| \frac{1}{2} \frac{\dot{X}}{X - 1} \right| =
\frac{6 H X}{3X - 1} ,
\label{eq040}
\end{equation}
$H = \dot{a}/a$ being the Hubble parameter, and the Thomas-Fermi condition (\ref{eq8}) becomes
\begin{equation}
\frac{6 H X}{3X - 1} \ll m \, .
\label{eq041}
\end{equation}
Note that, just as in the canonical model, the ghost condensation point $X_{c} = 1$ is reached asymptotically as $X - 1 \simeq \epsilon/a^{3} \ll 1$ according to (\ref{eq039}) with
$\epsilon^{- 1/3}$, a parameter giving an upper bound to the proximate redshifts.

In addition to (\ref{eq8}), one should check the Thomas-Fermi approximation against the full
energy density and pressure arising from (\ref{eq7}): 
\begin{equation}
\rho = \frac{m^{2}}{2} \, \phi^{2} \, (X + 1) + \frac{\lambda}{4} \, \phi^{4} + \frac{1}{2} \, \dot{\phi}^{2} \;\; ,
\label{eq19}
\end{equation}
\begin{equation}
p = \frac{m^{2}}{2} \, \phi^{2} \, (X - 1) + \frac{\lambda}{4} \, \phi^{4} + \frac{1}{2} \, \dot{\phi}^{2} \;\; .
\label{eq20}
\end{equation}
For this purpose we assume minimal coupling and ignore other forms of matter since we are interested in the vacuum state.

Using (\ref{eq038}), (\ref{eq040})
and the leading approximation to the Friedmann equation
\begin{equation}
H^{2} \simeq \frac{8 \pi G}{3} \, \rho_{TF} ,\; \;\; \rho_{TF} =
\frac{m^{4}}{4 \lambda} \, (3 X + 1) (X - 1)  \label{eq21}
\end{equation}
we obtain the following: near the condensation point (\ref{eq041}) becomes
\begin{equation}
\left[ 24 \pi   \left( \frac{G m^{2}}{\lambda} \right) \, (X - 1) \right]^{1/2} \ll 1
\label{eq22}
\end{equation}
while
\begin{equation}
\rho/\rho_{TF} \simeq 1 + 12 \pi \; \left( \frac{G m^{2}}{\lambda} \right) \, (X - 1)
\label{eq23}
\end{equation}
\begin{equation}
p/p_{TF} \simeq 1 + 48 \pi \; \left( \frac{G m^{2}}{\lambda} \right) 
\label{eq24}
\end{equation}
where $p_{TF} = {\cal{L}}_{GC}$ of (\ref{eq13}). 
Hence there are no large corrections to the Thomas-Fermi approximation.
 As $\rho_{TF} \simeq m^{4} \epsilon/\lambda a^{3}$ 
this conclusion is not qualitatively changed by the inclusion of ordinary matter. 
To gain some perspective, taking $\lambda$ of order unity, 
the limiting redshift at electron-positron annihilation
 $Z \simeq 1.7 \times 10^{10}$, and $\rho_{TF0}$ of the order of the critical density today,
 then $m \simeq$ 50 keV and for the worst case,  
the fractional correction in (\ref{eq24}) 
is $48 \pi \; G m^{2}/ \lambda \simeq 3 \times 10^{-45}$.

We now generalize these ideas to obtain an unholey gauged ghost condensate 
\cite{che13}, 
starting from the Abelian Higgs model\footnote{In a similar model discussed recently 
\cite{hash} the negative sign in front
of the kinetic term
was put in by hand.}
\begin{equation}
{\cal L}_{\rm H} = - \frac{1}{4 g^{2}}  F_{\mu \nu} 
F^{\mu \nu} + g^{\mu \nu} (D_{\mu} \Phi )^{*} (D_{\nu} \Phi ) - m^{2} | 
\Phi |^{2} - \lambda | \Phi |^{4}  , 
\label{eq25}
\end{equation}
with $F_{\mu \nu} = \partial_{\mu} A_{\nu} - \partial_{\nu} A_{\mu}$ and $D_{\mu} 
\Phi = \partial_{\mu} \Phi + i A_{\mu} \Phi$. Using the same decomposition of the 
field $\Phi$ as in (\ref{eq6}), and making the unitary gauge transformation
\begin{equation}
A_{\mu} \rightarrow A_{\mu} - m \theta_{,\mu}
\label{eq26}
\end{equation}
we arrive at
\begin{equation}
{\cal L}_{\rm H} = - \frac{1}{4 g^{2}}  F_{\mu \nu} 
F^{\mu \nu} + \frac{g^{\mu \nu}}{2} \left( \phi_{, \mu} \phi_{, \nu} + 
\phi^{2} A_{\mu} A_{\nu} \right) - \frac{1}{2} m^{2} \phi^{2} -
\frac{\lambda}{4} \phi^{4}   .
\label{eq27}
\end{equation}
Introducing again the Thomas-Fermi approximation (\ref{eq8}) yields
\begin{equation}
{\cal L}_{\rm HTF} = - \frac{1}{4 g^{2}}  F_{\mu \nu} 
F^{\mu \nu} +  \frac{\phi^{2}}{2} \left( A_{\mu} A^{\mu} - m^{2} \right)
- \frac{\lambda}{4} \phi^{4}  ,
\label{eq28}
\end{equation}
and eliminating $\phi$, the gauged ghost condensate is described by
\begin{equation}
{\cal L}_{\rm GGC} = - \frac{1}{4 g^{2}}  F_{\mu \nu} 
F^{\mu \nu} +  \frac{1}{4 \lambda} \left( A_{\mu} A^{\mu} - m^{2} \right)^{2} \Theta
\left(A_{\nu} A^{\nu} - m^{2} \right)  ,
\label{eq29}
\end{equation}
where $\Theta$ is the Heaviside step function.

Since we are dealing with a renormalizable field theory, loop effects 
can be incorporated by first calculating the effective lagrangian and 
subsequently applying the Thomas-Fermi approximation to this effective lagrangian. 
For the Abelian $\sigma$ model (\ref{eq5}) in Minkowski space-time
\begin{equation}
{\cal L}_{\sigma\rm eff} = Z (| \Phi |^{2} ) \eta^{\mu \nu} \Phi_{, \mu}^{*} \Phi_{, \nu} 
- U ( | \Phi |^{2})
\label{eq30}
\end{equation}
up to two derivatives. The Thomas-Fermi approximation is
\begin{equation}
{\cal L}_{\sigma \rm eff\, TF} = 
\frac{m^{2} \phi^{2}}{2} \: Z ( \frac{\phi^{2}}{2} ) X -
U ( \frac{\phi^{2}}{2} ) \equiv M^{4} P (X) .
\label{eq31}
\end{equation}
Note that (31)
is a Legandre transform between $P(X)$ and
$U(\phi^{2}/2)$.
Thus {\it every} purely kinetic $k$-essence \cite{arm14}
type model, such as ghost condensation,  can be viewed as 
the Thomas-Fermi approximation to some (perhaps nonrenormalizable)
 complex scalar field theory (for related ideas see \cite{mal15}).
In the Chaplygin gas model \cite{bil4}, for example,
$Z = 1$ and $U = m^{2} |\Phi^{2}| + A/(4 m^2 | \Phi|^{2})$.

The form of effective lagrangian describing the gauged ghost condensate, 
up to second derivatives, is fixed by general coordinate and gauge invariance:
\begin{eqnarray}
{\cal L}_{eff} = - \frac{\epsilon \left( | \Phi |^{2} \right)}
{4 g^{2}}  F_{\mu \nu} 
F^{\mu \nu} +  Z \left( | \Phi |^{2} \right)  g^{\mu \nu} 
\left( D_{\mu} \Phi \right)^{*} \left( D_{\nu} \Phi \right) \nonumber\\
-  U \left( | \Phi |^{2} \right) + \psi
\left( | \Phi |^{2} \right)  R \;\;   .
\label{eq32}
\end{eqnarray}
In the following discussion we restrict ourselves to the one-loop approximation.
 We also consider only scalar loops, thus avoiding the issues of quantum gravity,
 gauge fixing, and infrared divergences; also, then, the ungauged ghost 
condensate results are obtained by taking $A^{\mu}$ to be a pure gauge.

The calculation of the effective potential and $Z(|\Phi|^{2})$ follows standard 
functional methods \cite{itz16}.
Imposing the renormalization conditions 
$U'(0) = m^{2}$ and $U''(0) = 2 \lambda$, the effective potential is given by
\begin{eqnarray}
U(|\Phi|^2) = m^2 |\Phi|^2 + \lambda |\Phi|^4 
 +  \frac{1}{32 \pi^2}  
\Bigg[ \left( m^2 + 4 \lambda |\Phi|^2
\right)^2  \ln \left( 1 + 
\frac{4 \lambda |\Phi|^2}{m^2} \right) \nonumber\\
  -    4 m^2 \lambda |\Phi|^2 - 24 \lambda^2 |\Phi|^4 \Bigg] \;\; .
\label{eq33}
\end{eqnarray}
The function $Z(|\Phi |^{2} )$ is found to be
\begin{equation}
Z(| \Phi |^{2}) =
1  +  \frac{\lambda}{6 \pi^{2}}  
\frac{\lambda | \Phi |^{2}}{\left[ m^{2} + 4 \lambda | \Phi |^{2} \right]} .
\label{eq34}
\end{equation}
We note that, at one-scalar-loop order, $Z(| \Phi |^{2})$ is finite without the need for 
renormalization; further it is positive definite and bounded from above by
$\lambda/(24 \pi^2)$.

The situation regarding $\epsilon (|\Phi|^2)$ is more delicate. 
It obtains by evaluating scalar vacuum polarization with
$m^{2} + 4 \lambda | \Phi |^{2}$ as the effective scalar mass.
In terms of the unrenormalized gauge coupling $g_{0}$
\begin{equation}
\frac{\epsilon ( | \Phi |^{2})}{g^{2}} =
\frac{1}{g_{0}^{2}} + \frac{1}{3} 
{\int^{\Lambda}}  \frac{d^{4}  l_{E} }{( 2 \pi)^{4}} 
\frac{1}{\left[ l_{E}^{2} + m^{2} + 4 \lambda | \Phi |^{2} \right]^{2}}\, ,
\label{eq35}
\end{equation}
where $l_{E}$ is the Euclidean loop 4-momentum and an implicit 
gauge invariant cutoff has been imposed. 
Note that $\epsilon ( | \Phi |^{2} )/ g^{2}$ is positive definite, 
as expected on general grounds \cite{bun3}. 
It is also a decreasing function of $| \Phi |^{2}$, 
reflecting that the theory is not asymptotically free. 
Introducing the usual renormalized gauge coupling
\begin{equation}
\frac{1}{g^{2}} = \frac{1}{g_{0}^{2}} + \frac{1}{3} 
{\int^{\Lambda}}  \frac{d^{4} l_{E}}{(2 \pi)^{4}} 
\frac{1}{(l_{E}^{2} + m^{2} )^{2}}\, ,
\label{eq36}
\end{equation}
one has
\begin{equation}
\epsilon (| \Phi |^{2} ) = 1 - \frac{g^{2}}{48 \pi^{2}}  \ln 
\left( 1 + \frac{4 \lambda | \Phi |^{2}}{m^{2}} \right)   
\label{eq37}
\end{equation}
up to terms of order $g^{2} | \Phi |^{2} / \Lambda^{2}$ .
Naively, equation (\ref{eq37}) suggests $\epsilon( | \Phi |^{2})$ can become 
negative for sufficiently large $| \Phi |^{2}$. 
One recognizes this as the field space version of the Landau ghost, a
different shade altogether from those under discussion. 
Since the infinite cutoff limit cannot be taken without either $g^{2} = 0$ 
(triviality) or $g_{0}^{2} < 0$ (intrinsic
instability), resummation in (\ref{eq37}) assures $\epsilon ( | \Phi |^{2} ) > 0$,
and hence no violation of the NEC. Similar conclusions have been drawn in 
\cite{con9} for the Higgs-Yukawa model.

The last term in (\ref{eq32}) is of interest for two reasons.
First, aside from the Einstein-Hilbert term
$- R/2K$, where $K = 8 \pi G$ is the inverse of the reduced Planck mass squared,
 a non-minimal coupling of the form $\xi R | \Phi |^{2}$ to 
the Ricci scalar $R$ is necessary to absorb divergences arising from the
 coupling of the graviton $h_{\mu \nu} = g_{\mu \nu} - \eta_{\mu \nu}$ 
to the energy-momentum tensor,
$- \frac{1}{2} h^{\mu \nu} T_{\mu \nu}$, in loop diagrams \cite{cal}. 
The non-minimal coupling itself also appears in loops. 
At one-loop order, imposing the renormalization conditions $\psi(0) = -1/(2K)$
and $\psi'(0) = \xi$, 
a straightforward calculation yields
\begin{equation}
\psi(| \Phi |^{2} ) = - \frac{1}{2K} + \xi | \Phi |^{2} + ( \xi - \frac{1}{6} ) \,
\Bigg[ \frac{(m^{2} + 4 \lambda | \Phi |^{2})}{16 \pi^{2}} \, \ln \left( 1 + \frac{4 \lambda | \Phi |^{2}}{m^{2}} \right) -
\frac{\lambda|\Phi|^{2}}{4 \pi^{2}} \Bigg].
\label{eq38}
\end{equation}
One notes that for the so-called conformal coupling, $\xi = 1/6$, 
the one-loop contributions cancel. Even if the renormalized $\xi = 0$, loop corrections will induce a non-minimal coupling.

The second reason why the non-minimal coupling is of interest is that 
it has been suggested as an avenue for violating the NEC \cite{bar18}.
To investigate this possibility
in the context of ghost condensation we omit the loop corrections for simplicity, 
and add the Einstein-Hilbert plus non-minimal terms to (\ref{eq27}): 
\begin{equation}
{\cal L}_{HG} = {\cal L}_{H} - 
\frac{1}{2K}  R + \frac{\xi \phi^{2}}{2}  R  .
\label{eq39}
\end{equation}
One notes that for $K \xi \phi^{2} > 1$ the sign of the coefficient of 
$R$ is reversed, equivalent classically to a reversal of sign for 
$T_{\mu \nu}$ in the Einstein equations corresponding to a NEC violation.

Equation (\ref{eq39}) is written in the Jordan frame. By a conformal transformation
\begin{equation}
g^{\mu \nu} = ( 1 - K \xi \phi^{2} )  \tilde{g}^{\mu \nu}
\label{eq40}
\end{equation}
we obtain in the Einstein frame
\begin{equation}
\tilde{\cal {L}}_{HG} = \tilde{\cal {L}}_{H} - \frac{1}{2K}  \tilde{R}  ,
\label{eq41}
\end{equation}
\begin{eqnarray}
\tilde{\cal {L}}_{H} = - \frac{1}{4 g^{2}}  F_{\mu \nu}  F^{\mu \nu} +
\frac{\tilde{g}^{\mu \nu} \phi_{, \mu} \phi_{, \nu}}
{2 (1 - K \xi \phi^{2})} 
\left[ 1 + \frac{6 K \xi^{2} \phi^{2}}{(1 - K \xi \phi^{2} )} \right]\nonumber\\
+ \frac{\phi^{2}  A_{\mu}  A^{\mu}}{2 (1 - K \xi \phi^{2} )} -
\frac{1}{(1 - K \xi \phi^{2} )^{2}}  \left[
\frac{m^{2}}{2}  \phi^{2} + \frac{\lambda}{4}  \phi^{4} \right]  .
\label{eq42}
\end{eqnarray}
This complicated expression for $\tilde{\cal {L}}_{H}$ can be greatly simplified 
by a field and quartic coupling redefinition
\begin{equation}
\tilde{\phi}^{2} = \frac{\phi^{2}}{1 - K \xi \phi^{2}}\, , \hspace{1cm}
\tilde{\lambda} = \lambda + 2 K \xi m^{2}
\label{eq43}
\end{equation}
yielding
\begin{eqnarray}
\tilde{\cal {L}}_{H} = - \frac{1}{4 g^{2}}  F_{\mu \nu}  F^{\mu \nu} +
 \frac{1}{2}  \frac{(1 + 6 K \xi^{2} \tilde{\phi}^{2})}
 {(1 + K \xi \tilde{\phi}^{2})^{2}} 
  \tilde{g}^{\mu \nu} \tilde{\phi}_{\mu} \tilde{\phi}_{\nu}\nonumber\\
+ \frac{\tilde{\phi}^{2}}{2}  
(A_{\mu}  A^{\mu} - m^{2}) - \frac{\tilde{\lambda}}{4} 
\tilde{\phi}^{4}  .
\label{eq44}
\end{eqnarray}
Thus the basic effect of gravity is a slight change in the Thomas-Fermi condition
\begin{equation}
\frac{(1 + 6 K \xi^{2} \tilde{\phi}^{2})}
{(1 + K \xi \tilde{\phi}^{2})^{2}} \:
\tilde{g}^{\mu \nu} \: \frac{\tilde{\phi}_{, \mu}}{\tilde{\phi}}\, 
\frac{\tilde{\phi}_{\nu}}{\tilde{\phi}} \ll m^{2}  .
\label{eq45}
\end{equation}
In particular, there is no vestige of a NEC violation provided $\tilde{\phi}$ is
real. The would-be NEC violation recurs if $\tilde{\phi}$ is imaginary, 
as is evident from (\ref{eq43}) - in that case one should make the inversion
$\tilde{g}_{\mu \nu} \rightarrow - \tilde{g}_{\mu \nu}$.
Moreover, there is no regular way to join real $( \tilde{\phi}^{2} \geq 0)$ with imaginary $( \tilde{\phi}^{2} \leq - 1/K \xi)$ $\tilde{\phi}$,
which implies that any NEC violation is permanent
and disconnected from the ghost condensation point.
Thus a NEC violation, through imaginary $\tilde{\phi}$, is a true phantom.

A secondary effect of the original non-minimal coupling is that ordinary matter 
still sees the Jordan frame metric
\begin{equation}
g_{\mu \nu} = ( 1 + K \xi \tilde{\phi}^{2} )  \tilde{g}_{\mu \nu}  .
\label{eq46}
\end{equation} 
Hence, nonconformal matter will perturb the nontrivial solution
\begin{equation}
\tilde{\phi}^{2} = \frac{A_{\mu}  A^{\mu} - m^{2}}{\tilde{\lambda}}  ,
\label{eq47}
\end{equation}
and pick up a gravitational strength coupling to the condensate. 
The key point is that
these effects are not introduced {\it ad-hoc},
but determined by the underlying theory.

In summary, we have shown that, by applying the long-wavelength 
Thomas-Fermi approximation to the Abelian Higgs field theory, 
it is possible to construct a model that shares the essentials of a 
gauged ghost condensation, but does not exhibit the instabilities 
associated with the violation of the NEC.
It is worthwhile emphasizing that the Thomas-Fermi approximation effectively
 provides a specific UV completion for a specific set of ghost condensate models.
For example, the Lagrangian (\ref{eq13}) has (\ref{eq7}) as its UV completion. Similarly, 
the Lagrangian (\ref{eq29}) has (\ref{eq25}) as its UV completion.
This provides a theoretical underpinning to the class of 
modified graviton and spontaneous Lorentz violation models previously 
introduced in a purely phenomenological way in \cite{ark5} and \cite{che13}.

\subsection*{Acknowledgments}
This research is supported by the Foundation for Fundamental Research, 
the National Research Foundation of South Africa and the Ministry of 
Science and Technology of the Republic of Croatia 
under Contract No. 098-0982930-2864.

\end{document}